\documentstyle[prb,aps,epsf,psfig,multicol]{revtex}
\newcommand{\la}{\langle}
\newcommand{\ra}{\rangle}
\newcommand{\ua}{\uparrow}
\newcommand{\da}{\downarrow}

\newcommand{\rar}{\rightarrow}
\newcommand{\e}{\epsilon}
\newcommand{\om}{\omega}
\newcommand{\vk}{{\vec k}}

\newcommand{\vq}{{\vec q}}

\newcommand{\vp}{{\vec p}}
\newcommand{\vpp}{{\vec p\,'}}
\begin{document}
\draft

\title{Spin wave dispersion in La$_2$CuO$_4$}

\author{N. M. R. Peres and M. A. N. Ara\'ujo}
\address{Departamento de F\'{\i}sica, Universidade de \'Evora,
Rua Rom\~ao Ramalho, 59, P-7000-671 \'Evora, Portugal\\ and 
Centro de F\'{\i}sica da Universidade do Minho, Campus Gualtar, 
P-4700-320 Braga, Portugal}

\date{\today}

\maketitle

\begin{abstract}
We calculate the antiferromagnetic spin-wave dispersion in a 
half-filled Hubbard model for a two-dimensional square lattice, and find
it to be in excellent agreement with recent high-resolution inelastic 
neutron scattering performed on La$_2$CuO$_4$ 
[Phys. Rev. Lett. {\bf 86}, 5377 (2001)].

\end{abstract}
\vspace{0.3cm}
\pacs{PACS numbers: 75.30.Ds, 71.10.Fd, 75.40.Gb}
\begin{multicols}{2}

After over a decade of intense research on the microscopic origin of
high-temperature  superconductivity in cuprates, there is no general
consensus on the microscopic Hamiltonian suitable for describing 
these materials. Nevertheless, it appears that  magnetic
fluctuations must play an important role. 
Therefore, the study of  magnetic fluctuations in the high-temperature
superconductor parent compounds, such as La$_2$CuO$_4$, is an important
field of research, both theoretical and experimental.

 In two recent  papers, \cite{coldea,ronnow}
high-resolution inelastic neutron-scattering measurements 
were performed
on two different two-dimensional spin-1/2 quantum antiferromagnets. These
are 
copper deuteroformate tetradeuterate (CFTD) and La$_2$CuO$_4$. Surprisingly,
the dispersion at the zone boundary  
observed in the two materials,
does not agree with spin-wave theory predictions.\cite{igarashi} Moreover the 
amount of dispersion is not the same for both materials. In CFTD
the dispersion is about 6\% from $\omega(\pi/2,\pi/2)$ to 
$\omega(\pi,0)$, whereas in  La$_2$CuO$_4$ it is about -13\% along
the same direction. In the case of CFTD the dispersion at the zone boundary
can be explained using the nearest-neighbor Heisenberg model alone,
\cite{ronnow} and high-precision quantum Monte Carlo simulations
have confirmed  that this is so.\cite{sandvik} On the other hand, an
explanation for the observed dispersion in  La$_2$CuO$_4$  was proposed, 
\cite{coldea}  using   an extended Heisenberg model  
\cite{takahashi,macdonald} involving first-,
second-, and third-nearest-neighbor interactions
as well as interactions among four spins. This extended model
was obtained from the Hubbard model, using perturbation theory, and is
diagonalized afterward using classical (large-$S$) linear spin-wave theory. 
\cite{coldea}

The La$_2$CuO$_4$ results clearly show that the usual Heisenberg model
is insufficient to explain the experimental data, and that the Hubbard model
is the correct Hamiltonian for describing the magnetic interactions in the
cuprates. \cite{anderson} In this work we do not use perturbation
theory for deriving an effective magnetic Hamiltonian. Instead we
 work directly with the Hubbard model. 
We consider a half-filled Hubbard model in a spin-density-wave (SDW)-
broken 
symmetry ground state and, by summing up all  ladder diagrams, we
compute the transverse spin susceptibility and from this obtain the 
spin-wave dispersion. 

The Hubbard model for a square lattice of $N$ sites is defined as
\begin{eqnarray}
H&=&-t\sum_{\la i,j\ra,\sigma}(c^{\dag}_{i,\sigma}c_{j,\sigma}+
c^{\dag}_{j,\sigma}c_{i,\sigma})+\mu\sum_{i,\sigma}
c^{\dag}_{i,\sigma}c_{i,\sigma}\nonumber\\
&+&U\sum_i
c^{\dag}_{i,\ua}c_{i,\ua}c^{\dag}_{i,\da}c_{i,\da}\,,
\nonumber\\
&=&\sum_{\vk,\sigma}[\e(\vk)-\mu]c^{\dag}_{\vk,\sigma}c_{\vk,\sigma}+
H_U\,,
\label{hubbard}
\end{eqnarray}
where the sum over $\la i,j\ra$ counts each  pair of 
nearest neighbors only once, the momentum 
$\vk$ runs over  the Brillouin zone,   
the electronic energy dispersion $\e(\vk)=-2t\cos k_x-2t\cos k_y$ has 
the nesting vector $\vec Q=(\pi,\pi)$, and 
\begin{equation}
H_U=\frac U N\sum_{\vk,\vk',\vq}
c^{\dag}_{\vk,\ua}
c_{\vk-\vq,\ua}c^{\dag}_{\vk',\da}c_{\vk'+\vq,\da}\,.
\label{hu}
\end{equation} 

\begin{figure}[htf]
\begin{center}
\epsfxsize=7cm
\epsfbox{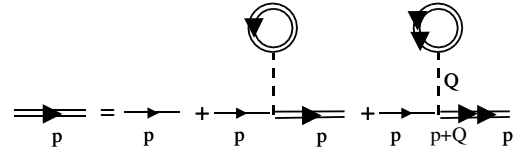}
\end{center}
\caption{Feynman diagrams corresponding to the SDW mean-field equation for 
$G_{\sigma}(\vp;\tau-\tau')$.}
\label{fdngf}
\end{figure}

\begin{figure}[ht]
\begin{center}
\epsfxsize=7cm
\epsfbox{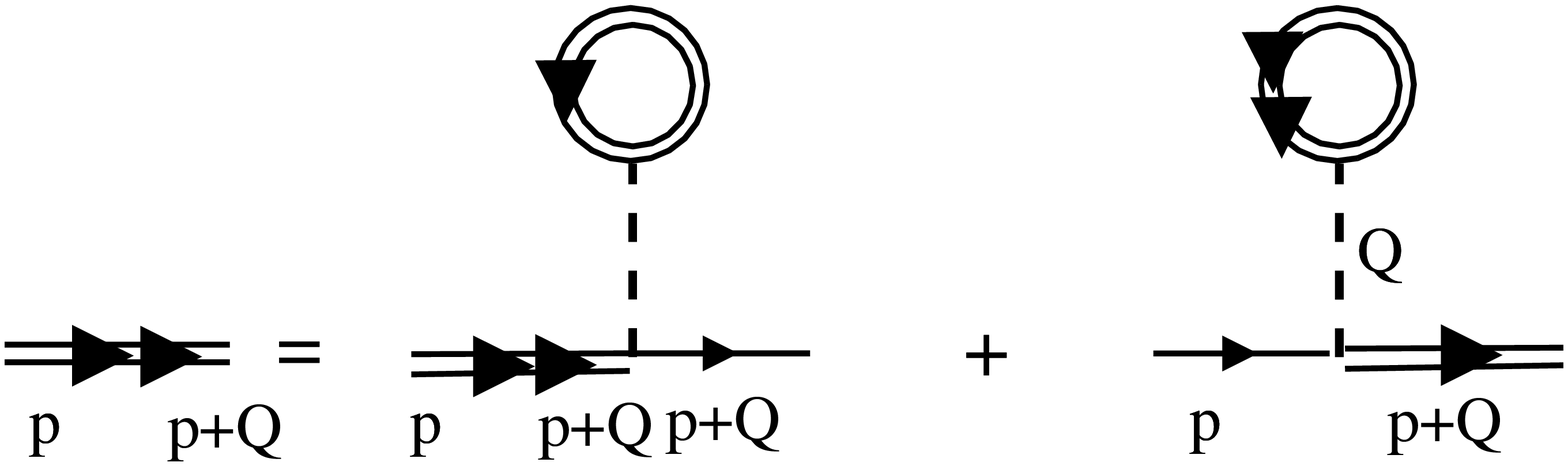}
\end{center}
\caption{Feynman diagrams corresponding to the SDW mean-field equation for 
$F_{\sigma}(\vp;\tau-\tau')$}
\label{fdagf}
\end{figure}

\begin{figure}[ht]
\begin{center}
\epsfxsize=7cm
\epsfbox{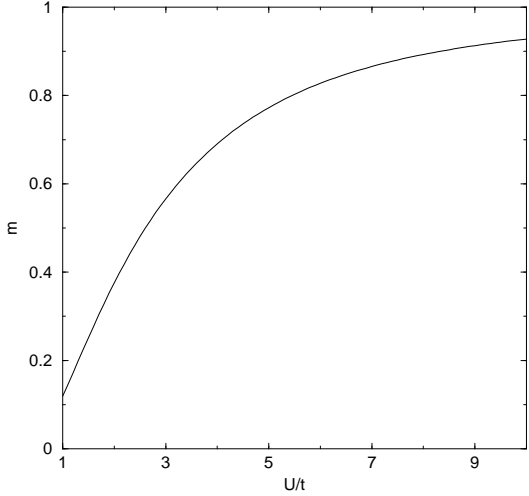}
\end{center}
\caption{Staggered magnetization as function of $U/t$ at $T=0$ K.}
\label{fstag}
\end{figure}

The broken-symmetry state is introduced by considering
the existence of an off-diagonal Green's function given by 
\begin{equation}
F_{\sigma}(\vp;\tau-\tau')=-\la T_{\tau}
c_{\vp\pm\vec Q,\sigma}(\tau)c^\dag_{\vp,\sigma}(\tau')\ra\,.
\end{equation}
in addition to the usual   Green's function: 
\begin{equation}
G_{\sigma}(\vp;\tau-\tau')=-\la T_{\tau}
c_{\vp,\sigma}(\tau)c^{\dag}_{\vp,\sigma}(\tau')\ra\,.
\end{equation} 
In mean-field theory for a SDW, the two propagators are related as shown
in the Feynman diagrams depicted in Figs.  \ref{fdngf} and \ref{fdagf}.
Single (doubled) arrowed lines represent diagonal (off-diagonal) propagators.
The single line represents the free propagator, whereas  double lines represent 
mean-field propagators. The Hubbard interaction is represented by 
a dashed line.
The solution to the mean-field equations yields the  
SDW staggered magnetic moment  $m$,
which is defined  as
\begin{equation}
\frac 1 N\sum_{\vp}\la
c^\dag_{\vp+\vec Q,\sigma}c_{\vp,\sigma}\ra=\frac m 2\sigma\,.
\end{equation}
The staggered magnetic moment
$m$ is reduced from its N\'eel value for finite values of $t/U$, and its
behavior as function of $t/U$ at zero temperature 
is shown in Fig. \ref{fstag}
Both the above Green's functions have  two-pole structures given by
\begin{eqnarray}
G_{\sigma}(\vp,i\om_n)&=&\frac {u_{\vp}}{i\om_n-E_+(\vp)}+
\frac {v_{\vp}}{i\om_n-E_-(\vp)}\,,\\
F_{\sigma}(\vp,i\om_n)&=&\frac {\tilde u_{\vp,\sigma}}{i\om_n-E_+(\vp)}+
\frac {\tilde v_{\vp,\sigma}}{i\om_n-E_-(\vp)}\,,
\end{eqnarray}
where the energies $E_{\pm}$ are given by
\begin{equation}
E_\pm(\vp)=\frac {\xi(\vp)+\xi(\vp+\vec Q)}2 +U\frac n2 \pm
E(\vp)\,,
\end{equation}
where $E(\vp)=\frac 1 2\sqrt{[\xi(\vp)-\xi(\vp+\vec Q)]^2+U^2m^2}$,
and
$\xi(\vp)=\e(\vk)-\mu$,
and the coherence factors read
\begin{equation}
u_{\vp}=\frac {E_+-\xi(\vp+\vec Q)-Un/2}{E_+-E_-}\,,
\hspace{0.5cm}
v_{\vp}=\frac {E_+-\xi(\vp)-Un/2}{E_+-E_-}\,,
\end{equation}
and
\begin{equation}
\tilde u_{\vp,\sigma}=\frac {Um\sigma/2}{E_+-E_-}\,,
\hspace{0.5cm}
\tilde v_{\vp,\sigma}=-\frac {Um\sigma/2}{E_+-E_-}\,.
\end{equation}

It is known that the mean-field treatment of the spin dynamics
of itinerant strongly correlated electronic systems
yields satisfactory results, as  
in the study of the dynamic spin response function 
of the cuprates'  superconducting state. \cite{lee}
In order to describe the spin dynamics of the system, we consider 
the  transverse spin susceptibility
$ \chi_{-+}(\vq,i\om_n)$,  which is defined
as  
\begin{equation}
\chi_{-+}(\vq,i\om_n)=\mu^2_B\int_0^\beta d\,\tau e^{i\om_n\tau}
\la T_\tau S^-(\vq,\tau)S^+(\vq,0)\ra\,,
\end{equation}
where $\beta=1/T$  is the inverse temperature,
$T_\tau$ is the chronological order operator (in imaginary time),
$S^-(\vq)= \sum_{\vp}c^\dag_{\vp,\da}c_{\vp+\vq,\ua}$,
and $S^+(\vq)=[S^-(\vq)]^\dag$.
The above expression can be written as
\begin{eqnarray}
\chi_{+-}(\vq,&i&\om_n)=\mu^2_B
\sum_{n=0}^\infty
\int_0^\beta d\,\tau\sum_{\vp,\vpp}e^{i\om_n\tau}
\la T_\tau [-\int_0^\beta d\,\bar\tau H_U(\bar\tau)]^n\nonumber\\
&&c^\dag_{\vp,\da}(\tau)c_{\vp+\vq,\ua}(\tau)
c^\dag_{\vpp+\vq,\ua}(0)c_{\vpp,\da}(0) \ra_{d.c.}\,,
\label{chi}
\end{eqnarray}
where $d.c.$ stands for differently connected diagrams.
\begin{figure}[ht]
\begin{center}
\epsfxsize=7cm
\epsfbox{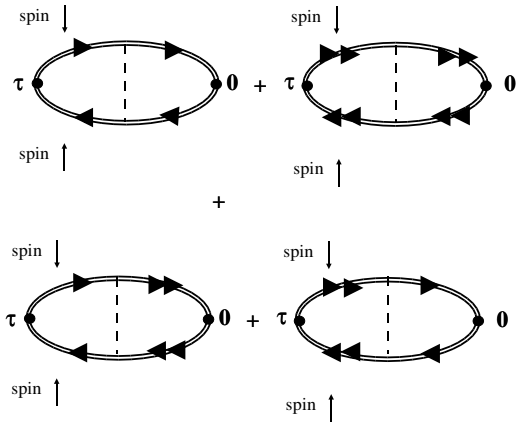}
\end{center}
\caption{The four first-order bubble
diagrams for $\chi^1_{+-}(\vq,i\om_n)$ in the SDW state. A summation over 
the internal momentum along the interaction line is implied.}
\label{c1a}
\end{figure}
We now compute $\chi_{-+}(\vq,i\om_n)$ by summing up all ladder diagrams
and taking into account the existence of two possible Green's-function lines.
The first-order  bubble diagrams are shown  in Fig. \ref{c1a}.
The complete ladder summation  is given by 
\begin{equation}
\chi_{+,-}^{ladder}(\vq,i\om_n)=
\frac{-(x+y)[1+\lambda(\bar x+y)]+\lambda(z_1+z_2)^2}{[1+\lambda (x+y)]
[1+\lambda(\bar x+y)]-\lambda^2(z_1+z_2)^2}\,,
\label{chiladder}
\end{equation}
with $\lambda=U/N$ and 
\begin{eqnarray}
x(\vq,i\om_n)&=&\frac {1}{\beta}\sum_{n\,',\vp}
G_\da(\vp,i\om_n\,')G_\ua(\vp+\vq,i\om_n\,'+i\om_n)\,,\nonumber\\
y(\vq,i\om_n)&=&\frac {1}{\beta}\sum_{n\,',\vp}
F_\da(\vp,i\om_n\,')F_\ua(\vp+\vq,i\om_n\,'+i\om_n)\,,\nonumber\\
z_1(\vq,i\om_n)&=&\frac {1}{\beta}\sum_{n\,',\vp}
G_\da(\vp,i\om_n\,')F_\ua(\vp+\vq,i\om_n\,'+i\om_n)\,,\nonumber\\
z_2(\vq,i\om_n)&=&\frac {1}{\beta}\sum_{n\,',\vp}
F_\da(\vp,i\om_n\,')G_\ua(\vp+\vq,i\om_n\,'+i\om_n)\,,\nonumber\\
\label{unit}
\bar x(\vq,i\om_n)&=&\frac {1}{\beta}\sum_{n\,',\vp}
G_\da(\vp,i\om_n\,')G_\ua(\vp+\vq+\vec Q,i\om_n\,'+i\om_n)\nonumber\,.
\end{eqnarray}
The  retarded susceptibility  is obtained from Eq. (\ref{chiladder}),
performing the analytical continuation $i\om_n\rar \omega + i0^+$. 
The poles of  the  retarded susceptibility
 give the energy $\om(\vq)$ of the  spin excitations
of the system as well as their lifetimes. The equation
for the poles is
\begin{eqnarray}
\{1&+&\lambda [x(\vq,\om)+y(\vq,\om)]\}\{
1+\lambda [\bar x(\vq,\om)+y(\vq,\om)]\}\nonumber\\
&-&\lambda^2[z_1(\vq,\om)+z_2(\vq,\om)]^2=0\,.
\label{modes}
\end{eqnarray} 
It is not possible to solve Eq. (\ref{modes})
 analytically for arbitrary  $U$ and $t$. 
In the limit  $t/U\rightarrow 0$  at half-filling,  from (\ref{modes})
we recover 
the same result as in  linear spin-wave theory for the nearest-neighbor
Heisenberg model,
\begin{equation}
\omega(q_x,q_y)\rightarrow\frac {4t^2}{Um}
\sqrt{4-(\cos q_x+\cos q_y)^2}\,,
\label{swlt}
\end{equation}
which predicts that $\omega(0,\pi)=\omega(\pi/2,\pi/2)$, in disagreement
with the experimental data,\cite{coldea} and $m\rightarrow 1$, showing no
magnetic moment reduction from the N\'eel state value.
Of course Eq. (\ref{swlt}) holds asymptotically for  $t/U\ll 1$. 
We remark that the factor $1/m$ plays
the role of the quantum renormalization factor $Z_c$.\cite{coldea}
\begin{figure}[f]
\begin{center}
\epsfxsize=8cm
\epsfbox{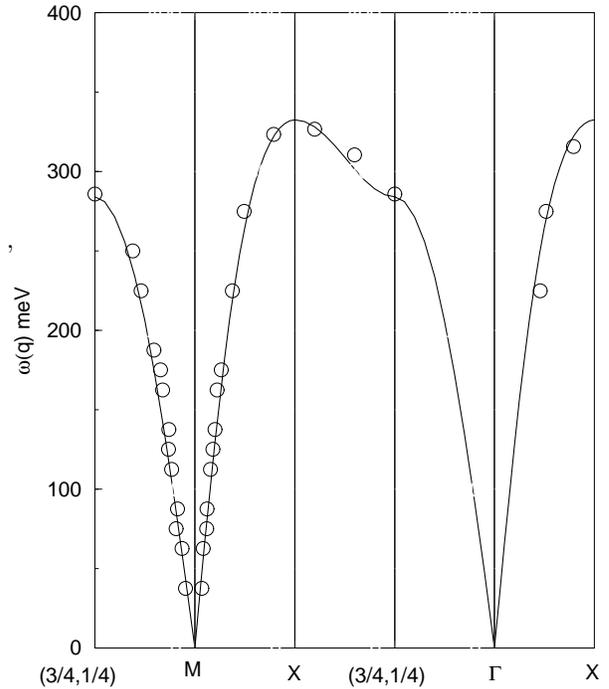}
\end{center}
\caption{Spin-wave dispersion, in meV, along 
high-symmetry directions in the Brillouin zone. 
The circles are the data reported in
Ref. \cite{coldea} at $10$ K. 
The solid line is the analytical result (at 0 K) for 
$U=1.8$ eV and $t=0.295$ eV. The momentum is in units of $2\pi$, and
$M=(1/2,1/2)$, $X=(1/2,0)$, and $\Gamma=(0,0)$.}
\label{fsw}
\end{figure}

On the other hand, for  finite  $t/U$,  Eq. (\ref{modes}) has to be solved
numerically. Considering the half-filled case
($\mu=0$), appropriate for
La$_2$CuO$_4$, we computed the spin-wave dispersion $\om(\vq)$ for
$U=1.8$ eV and $t=0.295$ eV along 
the high symmetry directions, in the two-dimensional
Brillouin zone, considered in Ref. \onlinecite{coldea}. These
values agree with those used in Ref. \onlinecite{coldea}:
$U=2.2\pm 0.4$ eV and $t=0.30\pm 0.02$  eV.
The results (solid line)
are given in Fig. \ref{fsw} together with the experimental results (in circles)
at $T$=10 K.
It is clear that $\omega(0,\pi)>\omega(\pi/2,\pi/2)$, that is, 
a dispersion at the zone boundary is obtained. For these values of
$t$ and $U$ the staggered magnetic moment is $m=0.832$, and therefore
$1/m=1.20$, which agrees well with $Z_c=1.18$ used to fit the data
in Ref. \onlinecite{coldea}.

In their interpretation of the experimental data, 
the authors of Ref. \onlinecite{coldea} considered an  effective Hamiltonian 
incorporating  ring exchange. In such a model
the electron  not only makes a virtual
trip to its nearest neighbor, but  also makes virtual excursions  around
a loop visiting its second neighbors. If we had written the transverse 
susceptibility in coordinate space, it would be clear that in a ladder
summation the electron goes around larger and larger rings before it
comes back to the original site with its spin flipped. 
Therefore, such a good
agreement between the perturbation theory calculation 
in Ref. \onlinecite{coldea} 
and our ladder summation  is  not  surprising.
 Althought we have not presented the results here, Eq. (\ref{modes}) also
predicts a continuum of excitations which are gapped relatively
to the ground state. 

In summary, we have shown that  a ladder
summation based on the SDW state can account satisfactorily 
 for the measured spin-wave dispersion
of  La$_2$CuO$_4$ at all energies. The quality of the fitting points out that
it is not needed to derive an effective spin Hamiltonian from the Hubbard
model in order to obtain agreement with the data. 

We gratefully acknowledge G. Aeppli for discussions, and for providing us
the experimental data we show in Fig. \ref{fsw}. We thank A. Sandvik
for calling our attention to this problem.

\end{multicols}


\begin{references}

\bibitem{coldea} 
R. Coldea, S. M. Hayden, G. Aeppli, T. G. Perring, C. D. Frost, T. E.
Mason, S.-W. Cheong, and Z. Fisk, Phys. Rev. Lett. {\bf 86}, 5377 (2001).
\bibitem{ronnow}
H. M. R{\o}nnow, D. F. McMorrow, R. Coldea, A. Harrison, I. D. Youngson, 
T. G. Perring, G. Aeppli, O. Sylju{\aa}sen, K. Lefmann, and C. Rischel,
Phys. Rev. Lett. {\bf 87}, 37202 (2001).
\bibitem{igarashi} J. Igarashi, Phys. Rev. B {\bf 46}, 10763 (1992).
\bibitem{sandvik} Anders Sandvik and Rajiv R. P. Sing, Phys. Rev. Lett.
{\bf 86}, 528 (2001).
\bibitem{takahashi}
M. Takahashi, J. Phys. C {\bf 10}, 1289 (1977).
\bibitem{macdonald}
A. H. MacDonald, S. M. Girvin, and D. Yoshioka, Phys. Rev. B {\bf 41},
2565 (1990); {\bf 37}, 9753 (1988).
\bibitem{anderson}
P. W. Anderson, Science {\bf 255}, 1196 (1987).
\bibitem{lee}
See, for example,  
Jan Brinckmann and Patrick A. Lee, 
cond-mat/0110316 [J. Low Temp. Physics (to be published)],
and references therein.

\end{references}
\end{document}